\documentclass[letterpaper, 10 pt, conference]{ieeeconf}  % Comment this line out if you need a4paper

\IEEEoverridecommandlockouts                              % This command is only needed if 
\usepackage{lipsum}
\usepackage{graphics} % for pdf, bitmapped graphics files
\usepackage{xcolor}
\usepackage{epsfig} % for postscript graphics files
\usepackage{times} % assumes new font selection scheme installed
\usepackage{amsmath} % assumes amsmath package installed
\usepackage{amssymb}  % assumes amsmath package installed
\usepackage{algorithm,algorithmic}
\usepackage{epsfig} % for postscript graphics files
\usepackage{caption}
\usepackage{subfigure}
\usepackage{bm}
\usepackage{lipsum}
\usepackage{comment}
\usepackage{xcolor}
\usepackage{relsize}
\usepackage{xurl}
\usepackage{booktabs}
\usepackage{multirow}
\usepackage{siunitx}
\usepackage{hyperref}
\newtheorem{theorem}{Theorem}

\usepackage{bm}

\usepackage{todonotes}

\newtheorem{definition}{Definition}
\newtheorem{assumption}{Assumption}
\newtheorem{proposition}{Proposition}
\newtheorem{corollary}{Corollary}

\newcommand{\param}{\theta}
\newcommand{\params}{\bm{\theta}}
\newcommand{\Param}{\Theta}
\newcommand{\Params}{\bm{\Theta}}

\newcommand{\strategy}{\gamma}
\newcommand{\strategies}{\bm{\gamma}}

\newcommand{\cstrategy}{\rho}
\newcommand{\cstrategies}{\bm{\rho}}

\DeclareMathOperator{\diag}{diag}

\title{\LARGE \bf
Smooth Games of Configuration in the Linear-Quadratic Setting
}

\author{Jesse Milzman$^{1}$, Jeffrey Mao$^{2}$, and Giuseppe Loianno$^{3}$% <-this % stops a space
\thanks{$^{1}$The author is with the DEVCOM Army Research Laboratory,
        2800 Powder Mill Rd, Adelphi, MD 20783, USA.
        {\tt\footnotesize  email: jesse.m.milzman.civ@army.mil.}}%
\thanks{$^{2}$The author is with the New York University, Tandon School of Engineering, Brooklyn, NY 11201, USA. {\tt\footnotesize email: jm7752@nyu.edu}.%
}
\thanks{$^{3}$The author is with the Department of Electrical Engineering and Computer Sciences, University of California, Berkeley, CA 94720, USA. {\tt\footnotesize email: loiannog@eecs.berkeley.edu}.}
}
\begin{document}

\maketitle
\thispagestyle{empty}
\pagestyle{empty}

%%%%%%%%%%%%%%%%%%%%%%%%%%%%%%%%%%%%%%%%%%%%%%%%%%%%%%%%%%%%%%%%%%%%%%%%%%%%%%%%
\begin{abstract}
Dynamic game theory offers a toolbox for formalizing and solving for both cooperative and non-cooperative strategies in multi-agent scenarios.
However, the optimal configuration of such games remains largely unexplored.
While there is existing literature on the parametrization of dynamic games, little research examines this parametrization from a strategic perspective—where each agent's configuration choice is influenced by the decisions of others.
In this work, we introduce the concept of a game of configuration, providing a framework for the strategic fine-tuning of differential games.
We define a game of configuration as a two-stage game within the setting of finite-horizon, affine-quadratic (AQ) differential games.
In the first stage, each player chooses their corresponding configuration parameter, which will impact their dynamics and costs in the second stage.
We provide the subgame perfect solution concept and a method for computing first stage cost gradients over the configuration space.
This then allows us to formulate a gradient-based method for searching for local solutions to the configuration game, as well as provide necessary conditions for equilibrium configurations over their downstream (second stage) trajectories.
We conclude by demonstrating the effectiveness of our approach in example AQ systems, both zero-sum and general-sum.
\end{abstract}

%%%%%%%%%%%%%%%%%%%%%%%%%%%%%%%%%%%%%%%%%%%%%%%%%%%%%%%%%%%%%%%%%%%%%%%%%%%%%%%%
\section{INTRODUCTION}

In the optimal control setting, it is well-known that fine-tuning the control system of an agent can improve its performance for a downstream task.
This can be formalized as the fine-tuning of physical parameters for the dynamics or cost equations governing a model-based controller in order to optimize the agent's objective --- either the intrinsic control costs from the parametrized system \cite{loquercio2022autotune, marco2016automatic, cheng2024difftune, romero2023weighted,configure_riveting} or an extrinsic reward separate from the controller itself \cite{oshin2022parameterized}.
However, in real-world scenarios, we often instead have multiple agents attempting to achieve disparate objectives, nonetheless forced to interact such as by sharing or competing for common resources or spaces.
Dynamic game theory offers a means of formalizing these scenarios, and finding their solutions.
However, there is little work considering game parametrization itself through the strategic lens, in which each agent will be fine-tuning their private degrees of freedom in light of similar considerations from other players. 
%However, the corresponding fine-tuning of dynamic games has been less studied.
%The parametrization of games has been studied in the context of incomplete information, sensitivity analysis of equilibrium solutions, and risk-aware control. 
%This is more like a related work thing

In this work, we introduce the concept of a game of configuration as a framework for the strategic, multi-agent fine-tuning of differential games.
We define games of configuration as two-stage games within the setting of finite-horizon, Linear-Quadratic (LQ) games --- an analytically well-understood starting point for differential games.
This choice is motivated by the successful application of iterative LQ solvers to find approximate solutions to more general systems, in both the single- and multi-agent settings \cite{fridovich2020efficient}.
In the first stage of a game of configuration, every player will have the opportunity to choose their private component $\param^i$ of the system's parameter vector $\params$, corresponding to preconfigurable degrees of freedom in their control capabilities and costs.
Then, the second stage is given as a parametrized differential (sub)game, in which every player pursues their feedback Nash strategy $\strategy(t,x \, ; \, \params)$ with complete information of the parameters $\params$.
After defining this notion and its solution concepts, we derive the gradient of the game's value function in order to develop an iterated best response (IBR)-based method for searching for local solutions.
Finally, we exhibit the value of both our framework and the gradient-based solver in both zero-sum and general-sum differential games.
Our technical contributions may be summarized as follows:
\begin{itemize}
    \item We propose a novel paradigm for configuration games to address the problem of non-cooperative, multi-agent preoconfiguration in differential games. To our knowledge, our approach is the first to this problem.
    \item We provide an analytic analysis of configuration games in the affine-quadratic (AQ) setting in order to find the gradient of the game's value as the solution to a linear ODE.
    We derive necessary conditions for equilibria over the configured trajectory, thereby constructively proving a special case of the envelope theorem for differential game theory 
    \cite{caputo2007envelope}.
    \item We use the gradient to implement an IBR-based solver for local solutions to configuration games. We demonstrate our approach on real-world inspired examples, including both a zero-sum and general-sum game.
    We also show the importance of our framework by demonstrating the suboptimality of a non-game-theoretic configuration baseline.
\end{itemize}

 \section{RELATED WORK}

Many problems of interest in mult-agent systems are modeled using the framework of dynamics games.
In multi-agent planning, for instance, agents non-cooperatively negotiate a common state, such as merging lanes in traffic or reaching their individual goals without collisions \cite{risk_aware_planning_wang,williams2023distributed}.
% \risk aware Wang - Lane merging without crashing the common resource is the road
% williams2023distributed - Each Agent needs to reach the goal. They need to negotiate their shared space so they don't crash with each other
\cite{risk_aware_planning_wang,williams2023distributed}.
In zero-sum scenarios, such as in pursuit evasion games \cite{GLIZER201522}, agents instead are at cross purposes and must act optimally while expecting their opponent to do likewise.
% Some examples including multi-agent planning where agents either need to negotiate a common resource cooperatively \cite{risk_aware_planning_wang, williams2023distributed} or in a zero sum case for example pursuit and evasion.
These multi-agent scenarios often have configurable parameters.
 In \cite{risk_aware_planning_wang} each agent has its own parameter affecting collision risk tolerance. The pursuit evasion game in \cite{GLIZER201522} has adjustable regularizers on both the pursuer and evader's acceleration action in its formulation. 
%Unsure about this example for configurable robotos to show dynamics can change or if we want to go into this
% Additionally besides optimization objectives, there is a class of robots \cite{WANG20237858, layout_design_multi_robot_transport} which can self configure themselves to solve tasks.
%%% Wang's work involves coalitions, and I don't really want to touch coalition formation if I can avoid it.... it's a very different kind of decision-space
Beyond tuning controllers, a more physical example may be found in the robotics literature concerned with the optimal spatial configurations for teams of robots \cite{layout_design_multi_robot_transport}.
% \footnote{\jesse{Jeff Twigg might have a paper here...}} 
For example, \cite{ layout_design_multi_robot_transport} involves first creating a rigid UAV formation for their downstream task of payload transportation \cite{layout_design_multi_robot_transport}.
This initial choice of configuration is smooth and creates different cooperative dynamics.
%% I'm happy with the first paragraph above for now. Still working my way through the rest.

% \vspace{-6pt}
Configuration and parameter-tuning are well-studied in robotics for the single-agent case.
% Common methods for tuning the configuration of a problem in robotics have focused on a single agent case.
These methods frequently involve objective-tuning for the controller, which has an intrinsic objective following a reference trajectory, based upon on some extrinsic objective function \cite{loquercio2022autotune, marco2016automatic, cheng2024difftune, romero2023weighted}.
Optimization is achieved through both gradient-based \cite{cheng2024difftune, romero2023weighted} and Bayesian methods \cite{loquercio2022autotune, marco2016automatic}. 
Similar approaches can be applied to the dynamic configuration of a robot manipulator, as in \cite{configure_riveting}, wherein first the optimal joint pose must be decided before the robot proceeds to the downstream task of riveting.
Most similar to our formulation, there is recent work that aims to optimize time-invariant configuration parameters directly for the underlying control problem, i.e. for the intrinsic control objective. Oshin \textit{et al.} proposed parameterized differential dynamic programming (PDDP) to jointly optimize invariant parameters and open-loop controls for a single-agent system, via a joint Newton's step in iterative LQ approximations \cite{oshin2022parameterized}.
Alternatively \cite{layout_design_multi_robot_transport} presents a two stage approach for preconfiguring a multi-drone formation for transporting a payload along a reference trajectory.
% First, a centralized planner solves an optimal drone formation, then the formation  carries a payload along a reference trajectory in the second.
Our work is motivated by this literature, but departs in considering the multi-agent case rather than the single-agent.

In the multi-agent setting, framing the configuration problem as a game is a natural perspective.
However, adding an initial stage to the game to allow for the strategic selection of parameters rarely appears in the literature.
Most examples are for specific applications rather than for a general case.
For example, \cite{ZHOU2022105376} uses game theory to model interactivity as a smooth continuous value between agents  before excuting formation control. 
\cite{WANG20237858} uses game theory to have agents discreetly choose roles before 
being assigned to a downstream attack defend set.
\cite{fisac2019hierarchical} uses game theory for autonomous driving where agents play a game to determine a value function then performs a short horizon planning with the previous game's value function as guidance. 

 \section{PROBLEM FORMULATION}
\label{sec:formulation}

In this section, we will describe the notion of a game of configuration within the LQ setting.
Much of our formulation could be easily be generalized to arbitrary differential games, and so we will briefly set the stage in that setting.
However, insofar as there are no general methods for solving the Hamilton-Jacobi-Isaacs (HJI) equation, it would also be far from trivial to solve for optimal configurations over arbitrary HJI solutions.
Therefore, we will ultimately restrict our study to the LQ setting, in which feedback Nash strategies are well-known to be provided by the solutions to a derived Riccati Differential Equation (RDE).

Consider an $N$-player differential game over time horizon $[0,T]$.
Traditionally, such games take the form
\begin{gather}
    \dot{x} = f(t, x, \bm{u}) ; \quad x(0) = x_0\\
    J^i(\bm{u}) = \int_{0}^T g^i(t,x,\bm{u}) dt + g^i_f(x(T),\bm{u}(T))
\end{gather}
where $\bm{u}(t) = (u^1, ..., u^N)(t)$, and each player is attempting to choose the control $u^i(t)$ that minimizes their total cost $J^i$.
% Strategies may have open-loop or feedback feedback information structure, i.e. $\strategy^i(t)$ or $\strategy^i(t,x)$.
We assume strategies have a feedback structure, i.e. $u^i(t) = \strategy^i(t,x)$.
% \footnote{In differential games, feedback strategies associated to value functionals satisfying general-sum HJI equations (\textcolor{red}{REF}) are Nash equilibria under both feedback and the stronger closed-loop perfect state (that is, full-memory) information structures. This holds for the feedback strategies in this paper, as well, although we will not discuss the distinction.} strategies $\gamma^i$ --- that is, strategies of the form $u^i(t) = \gamma^i(t)$ versus $u^i(t) = \gamma^i(t,x)$, respectively --- may be desired, if we ignore the possibility of imperfect information.
We overload our cost function to denote the value of a strategy vector, i.e. $J^i(\bm{\gamma}) = J^i(\bm{u})$ when $\bm{u}(t)$ is the $\bm{\gamma}$-induced open-loop representation of player controls.

The idea of a game of configuration is to extend these games to a two-stage process.
We assume that the second stage will be a differential game as above, except that both $f$ and $g^i$ will depend on a vector of parameters $\bm{\theta} = (\theta^1, ..., \theta^N)$.
The first stage, preceding this $\bm{\theta}$-specific dynamic game, will be one in which every player P$i$ will be choosing their parameter $\theta^i \in \Theta^i$ from a compact set $\Theta^i$.
This models a scenario in which every player is not merely planning for the dynamic game by choosing a dynamic strategy $\gamma^i$, but additionally has the opportunity to tune their private dynamics and costs to be more favorable.
For instance, in standard vehicles, tire pressure must be preconfigured, and there can be trade-offs between speed and control, especially when driving on diverse terrain.
% In industrial... \jesse{talk about Cournot competition, maybe??}

\subsection{Parametrized Affine-Quadratic Games}
\label{subsec:LQ_games}

We will concretize the notion of a game of configuration in the LQ setting, specifically in affine-quadratic (AQ) games.
An affine-quadratic game is a mild generalization of the traditional LQ game, in which state dynamics are allowed to include a $0$-th order driving term $c(t)$.
Our system's dynamics and player costs are given by
\begin{gather}
    \label{eq:LQ.dynamics}
    \dot{x} = A(t) x + \sum_{i=1}^N B^i(t ; \theta^i) u^i(t) + c(t), \quad x(0) = x_0, \\
   \nonumber
    J^i(\bm{u} | \params) = \frac{1}{2} \left[ \int_{0}^T \left\| x \right\|_{Q^i(t;\bm{\theta})}^2 + \sum_{j=1}^N \left\| u^j \right\|_{R^{ij}(t)} dt \right.\\
    \label{eq:LQ.cost}
    + \left. \left\| x(T) \right\|_{Q_f^i} \right],
\end{gather}
where $\left\| v \right\|_{M}^2 = v^\top M v$, state and controls have dimensions $x(t) \in \mathbb{R}^n$ and $u^i(t) \in \mathbb{R}^{m_i}$, and all matrices have the correspondingly appropriate shape.
We make a few notes.
First, we can accommodate 1st-order terms on the state in the cost, e.g. $x^\top Q x + \xi^\top x + \omega$, via completing the square and a simple change of variables in the state, neglecting any strategy-independent constant terms.
This is done in our general sum experiments described in Section~\ref{subsec:general_sum_form} specifically Eq.~\ref{eqn:general_cost}.
Second, when $c(t) = 0$ for all $t \in [0,T]$, we refer to the game as \textbf{linear-quadratic (LQ)}.
Third, in the two-player, zero-sum linear-quadratic case, we typically just represent the value of the game $J \triangleq J^1 = -J^2$ as a single equation, the minimizer's (P$1$'s) cost:
\begin{gather}
    \nonumber J(\bm{u} | \params) = \frac{1}{2} \left[ \int_{0}^T \left\| x \right\|_{Q(t;\bm{\theta})}^2 + \left\| u^1 \right\|^2 - \left\| u^2 \right\|^2  dt \right. \\
    \label{eq:LQ.cost.zerosum}
    + \left. \left\| x(T) \right\|_{Q_f} \right]
\end{gather}

\begin{theorem}[{\cite[Corr. 6.5]{bacsar1998dynamic}}]
\label{thm:basar.AQ_riccatti}
Let $\bm{\theta}$ be fixed for the \textbf{Affine-Quadratic (AQ)} differential game given by (\ref{eq:LQ.dynamics}-\ref{eq:LQ.cost}), and assume $Q^i(t;\bm{\theta}),Q_f^i,R^{ij} \geq 0$, and $R^{ii}(t) > 0$ for all $t,i \neq j$.
Then, if there exist solutions $P^i(t) = P^i(t;\bm{\theta}), \; t \in [0,T]$ for the coupled Riccati matrix differential equations:
\begin{gather}
    \label{eq:AQ_riccati}
    \dot{P}^i + P^i \tilde{F} + \tilde{F}^\top P^i + Q^i + \sum P^j S^{ij} P^j = 0; \; P(T) = Q_f^i, \\ 
    \label{eq:Sij}
    S^{ij}(t) = (B^j R^{jj-1} R^{ij} R^{jj-1} B^{j \top})(t), \\
    \label{eq:Ftilde}
    \tilde{F}(t) \triangleq  A(t) - \sum_{i=1}^N S^{ii}(t) P^i(t),
\end{gather}
then there exists a feedback Nash equilibrium to the game, 
% valid for closed-loop and memoryless perfect-state (CLPS and MPS) information structures as well, 
given by the strategies
\begin{equation}
\label{eq:AQ_riccati_strategy}
\gamma^{i\star}(t,x) = - R^{ii -1}(t) B^i(t)^\top (P^i(t) x(t) + \zeta^i(t)),    
\end{equation}
where $\zeta^i(\cdot) = \zeta^i(\cdot | \bm{\theta}) $ is the solution to the linear equation:
\begin{gather}
    \label{eq:zeta}
    \dot{\zeta}^i + \left[ \tilde{F}^\top \zeta^i + \sum_{j=1}^N P^j S^{ij} \zeta^j \right] + P^i \beta = 0 ; \quad \zeta^i(T) = 0, \\
    \beta(t) \triangleq c(t) - \sum_{i=1}^N S^{ii}(t) \zeta^i(t).
\end{gather}
For a given initial condition $x_0$, the value of the game is given by
\begin{equation}
    \label{eq:AQ_riccati_cost}
    J^i(\bm{\gamma^{\star}}; \bm{\theta} ) = \frac{1}{2} x_0^\top P^i(0) x_0 + \zeta^{i \top}(0) x_0 + \eta(0),
\end{equation}
where $\eta^i(\cdot) = \eta^i(\cdot | \bm{\theta})$ is defined as the solution to
\begin{equation}
    \label{eq:eta}
    \dot{\eta}^i + \beta^{\top} \zeta^i + \frac{1}{2} \sum_{j=1}^N \zeta^j S_{ij} \zeta^j = 0; \quad \eta^i(T) = 0.
\end{equation}
Note that, for \textbf{linear-quadratic (LQ)} games, i.e. when $c(t) = 0$ for all $t\in[0,T]$, we have that $\zeta^i \equiv 0$ and $\eta^i \equiv 0$.
\end{theorem}

We will develop our method for finding local solutions to configuration games in this general-sum, affine-quadratic context, as that will allow us the greatest flexibility in the systems that we may model or approximate.
In the case of a two-player, zero-sum linear-quadratic game (i.e. $N=2, J^1 = - J^2$, $R^{ij} = (-1)^{i-j} I$, and $c \equiv 0$), the solution to the game depends on the existence of a solution to single, simpler Riccati equation (see \cite[Thm. 6.17]{bacsar1998dynamic}):

\begin{corollary}
    A zero-sum, linear-quadratic game of finite horizon satisfying the conditions of Theorem~\ref{thm:basar.AQ_riccatti} has a feedback Nash equilibrium if there exists a solution $P(t), t \in [0,T]$ to the Riccati matrix differential equation:
    \begin{equation}
        \label{eq:riccati_zerosum}
        \dot{P} + PA + A^\top P + Q + P \tilde{S} P = 0; \quad P(T) = Q_f
    \end{equation}
    where $P$ now encodes P1's cost (the minimizer) and and P2's reward (the maximizer).
    The feedback Nash strategies $\strategies^{\star}$ are given by
    \begin{equation}
        \strategy^{i\star}(t,x) = (-1)^i B^i(t)^\top P(t) x(t)
    \end{equation}
    and the value of the game for $x(0) = x_0$ is given by
    \begin{equation}
        \label{eq:zerosum_LQ_game_value}
        J(\gamma^{\star}; \params) = \frac{1}{2} x_0^\top P(0) x_0
    \end{equation}
\end{corollary}
% \todo[inline]{Zero-sum version, if we have room}

\subsection{Games of Configuration in the Affine-Quadratic Setting}
\label{subsec:LQ_games_of_config}

Now that we have discussed how to solve affine-quadratic games with known parameters, we may take the next step and address the configuration problem.
In this section, we will define a game of configuration over a family of $\bm{\theta}$-parametrized affine-quadratic games.
We imagine this game as involving the strategic pre-configuration of a game's dynamics and costs, within the degrees of freedom afforded to that player to prepare.

The general idea is as follows: we construct a two-stage game in which we nest the $\params$-parametrized game from Sec.~\ref{subsec:LQ_games} as the second stage. Then, every player P$i$'s choice of $\param^i$ forms the first stage.
Temporally, the game proceeds as follows.
\begin{enumerate}[leftmargin=1.5cm, labelwidth=2.5cm]
    \item[\textbf{Stage 1.}] Every player P$i$ chooses $\param^i$ from a set $\Param^i$, simultaneously, fixing $\params$.
    \item[\textbf{Stage 2.}] In the $\params$-parametrized variant of the affine-quadratic game (\ref{eq:LQ.dynamics}-\ref{eq:LQ.cost}), every player chooses and pursues an feedback strategy $\strategy^i(\cdot, \cdot; \params)$, leading to final game cost $J^i(\strategies ; \params)$
\end{enumerate}

We can now formally define the game.
We will briefly ignore questions of regularity (or integrability) of any of the matrix coefficients under consideration.
\begin{definition}
\label{defn:game_of_config}
    An $N$-player game of configuration in the (finite-horizon, differential) affine-quadratic setting, which we also refer to as an \textbf{AQ game of configuration}, is given by the following components
    \begin{enumerate}
        \item To each player $i$, there is associated a closed interval $\Param^i = (\param^i_{\min}, \param^i_{\max}) \subset \mathbb{R}$.
        \item A time horizon $[0,T]$ for some finite $T>0$.
        \item The following matrix-valued functions defined on the product of $[0,T]$, $\Params^i$, and/or $\Params$ as appropriate:
        $A(t),B^i(t;\param^i), Q^i(t;\params),$ and $ R^{ij}(t)$ for all $i,j \in [N]$.
         % as in (\ref{eq:LQ.dynamics}-\ref{eq:LQ.cost}), with matrix-valued functions defined on the product of $[0,T]$, $\params^i$, and/or $\Params$ as appropriate.
        \item Affine dynamics vector $c(t)$ defined on $[0,T]$.
        \item Fixed terminal cost matrix $Q^i_f$ for each $i \in [N]$.
    \end{enumerate}
    For each player $i \in [N]$, a \textbf{strategy} to the game is given by the tuple $\cstrategy^i = (\param^i, \strategy^i)$, where $\param^i \in \Param^i$ is the (Stage 1) choice of parameter and $\strategy \in [0,T] \times \mathbb{R}^{n} \times \Params \to \mathbb{R}^{m_i}$ provides a Stage 2 feedback strategy $\strategy(\cdot, \cdot ; \params)$ for each $\params \in \Params$.
    We use $\cstrategies, \params,$ and $ \strategies$ to denote the concatenation of these strategy variables over players.
    The full set of player strategies $\cstrategies = (\params, \strategies)$ incurs the following total game cost for player $i$:
    \begin{equation}
        J^i(\cstrategies) \triangleq J^i(\strategies(\cdot \,;\, \params) \;|\; \params )
    \end{equation}
    where the right-hand side is defined as in (\ref{eq:LQ.dynamics}-\ref{eq:LQ.cost}).
    In the event that $c(t) = 0$ for all $t\in[0,T]$, i.e. the Stage 2 game is LQ, we refer to the full game as an \textbf{LQ game of configuration}.
    % We henceforce omit $\params$ from $\strategies(\cdot) = \strategies(\cdot \,;\, \params)$ when clear from context.
\end{definition}

Note that our definition of strategy requires players to specify a Stage 2 AQ strategy for non-visited branches of the game, i.e. for other possible choices of $\params' \in \Params$.
If we only required a fixed substrategy for the second stage, i.e. some $\gamma^i: [0,T] \times \mathbb{R}^n \to \mathbb{R}^{m^i}$ to be played no matter the Stage 1 choice, this would more easily allow for defections that would destabilize any potential equilibrium.
Moreover, since we're assuming feedback information structures, the only reason to use such a strategy structure would be if we did not expect agents to be able to observe or infer others' $\params^{-i}$ at the start of Stage 2, and adjust accordingly.
\begin{definition}
\label{defn:config_nash}
    Let an AQ game of configuration (Def~\ref{defn:game_of_config}) be given.
    We consider a strategy vector $\cstrategies^{\star} = (\params^{\star},\strategies^{\star})$ as a subgame perfect Nash equilibrium (SPNE) if the following two conditions hold for every $i \in [N]$:
    % \begin{enumerate}
    %     \item $J^i(\cstrategies^{\star}) \leq J^i(\cstrategy^i, \cstrategies^{-i \star})$
    % \end{enumerate}
    \begin{gather}
        \label{eq:config_nash.stage1}
        J^i(\cstrategies^{\star}) \leq J^i(\cstrategy^i, \cstrategies^{-i \star}), \quad \forall \cstrategy^i = (\param^i,\strategy^i) \\
        \nonumber
        J^i(\strategies^{\star}(\cdot \,;\, \params) \;|\; \params^{} ) \leq  \\
        \label{eq:config_nash.stage2}
        J^i \left( \strategy^i (\cdot \,;\, \params) ,\strategies^{-i \star}(\cdot \, ; \, \params) \; | \; \params \right), \quad
        \forall \gamma^i, \forall \params \in \Params
    \end{gather}
\end{definition}
Observe that (\ref{eq:config_nash.stage1}) merely requires a strategy to be optimal from the root of the game tree, i.e. at Stage 1.
However, with (\ref{eq:config_nash.stage2}) we assume sequential rationality on the part of every player, and assume they will pursue their subgame interests in Stage 2 no matter which $\params$ is collectively chosen by every player.

Up until this point, we have side-stepped the question of the properties we are assuming in Def.~(\ref{defn:game_of_config}).
Clearly, without some strong assumptions, we do not know if costs $J^i$ are even computable via (\ref{eq:LQ.dynamics}-\ref{eq:LQ.cost}), let alone if equilibria exist.
For all games of configuration under study in this work, we take the following on assumption.
\begin{assumption}
\label{assumption.game_config}
An AQ game of configuration satisfies this assumption if the following properties hold:
   \begin{enumerate}
       \item $A(t), R^{i,j}(t), c(t)$ are continuous on $[0,T]$. Moreover, for all $i,j$, $R^{ij}(t)$ is uniformly bounded above and below for all $t$ in the sense of $\alpha_1 \leq \lambda_{\min}(A(t))$ and $\left\|A(t) \right\|_{\mathcal{F}} < \alpha_2$ for some $\alpha_2 > \alpha_1 > 0$.
       % \item For every $\params \in \Params$ and $i \in [N]$, the matrices $B^i(t;\param^i), Q^i(t;\params)$ are continuous with respect to $t \in [0,T]$.
       % \item For each $i\in [N]$ and $t \in [0,T]$, $B^i(t; \param^i)$ is continuously differentiable with respect to $\param^i$ on $\Param^i$.
       % \item For each $i\in [N]$ and $t \in [0,T]$, $Q^i(t; \params)$ is continuously differentiable with respect to $\params$ over $\Params$.
       \item For each $i\in [N]$, $B^i(t; \param^i)$ is continuously differentiable with respect to $(t,\param^i)$ on $[0,T] \times \Param^i$.
       \item For each $i\in [N]$, $Q^i(t; \params)$ is continuously differentiable with respect to $(t,\params)$ over $[0,T] \times \Params$.
       \item For every $\params \in \Params$, there exist finite solutions $P^i_{\params}(t), t \in [0,T]$ to the coupled Riccati equation (\ref{eq:AQ_riccati}-\ref{eq:Ftilde}) for every $i \in [N]$. These, in turn, induce solutions $\zeta^i_{\params}$ and $\eta^i_{\params}$ to (\ref{eq:zeta}) and (\ref{eq:eta}), respectively.
       Moreover, these solutions are Lipschitz continuous on $[0,T] \times \Params$.
       % Moreover, there exists some $M>0$ such that the Frobenius norm $\left\| P^i_{\params} \right\|^2_{\mathcal{F}} \leq M$ for all $i \in [N], \params \in \Params, t \in [0,T]$.
   \end{enumerate} 
\end{assumption}
As we will see, these assumptions are sufficient for the properties we wish to demonstrate in the next section (Sec.~\ref{sec:approach}).
They are by no means sharp, but meet our needs in this work.
Using the Riccati-induced Nash for every $\params$-specific Stage 2 subgame, via Theorem~\ref{thm:basar.AQ_riccatti}, we can simplify our search for an SPNE to coupled optimization problems over the Riccati cost functions.
\begin{proposition}
\label{prop:stage1_nash}
    Let an AQ game of configuration satisfying Asm.~\ref{assumption.game_config} be given.
    For each player $i$ and $\params \in \Params$, let $\strategy^{i\star}_{\params}$ be the Riccati feedback strategy given by (\ref{eq:AQ_riccati_strategy}), and $\strategies_{\params} = (\strategy^i_{\params})_i$.
    For each $i \in [N]$, define the cost function
    \begin{equation}
        \label{eq:param_cost}
        J^i(\params) \triangleq J^i(\strategies^{\star}_{\params} \; | \; \params)
    \end{equation}
    over $\Params$, where we know the right-hand side takes the form as in (\ref{eq:AQ_riccati_cost})
    Then there exists a SPNE to the game of configuration if there exists a $\params^{\star} \in \Params$ such that
    \begin{equation}
        \label{eq:param_cost_Nash}
        J^i(\params^{\star}) \leq J^i(\param^i, \params^{-i \star}), \quad \forall \param^i \in \Param^i.
    \end{equation}
    The SPNE is given by $\cstrategies^{\star} = (\params^\star, \strategies^{\star})$, where $\strategy^{i\star}(\cdot \, ; \, \params) = \strategy^{i \star}_{\params}$ for all $i\in [N], \params \in \Params$.
\end{proposition}

% \begin{proposition}
% Let an AQ game of configuration satisfying Assmpt~\ref{assumption.game_config} be given, and assume there exists a SPNE $\cstrategies^{\star}=(\params^{\star},\strategies^{\star})$ to the game.
% For each player $i$ and $\params \in \Params$, let $\strategy^{i}_{\params}$ be the Riccati feedback strategy given by (\ref{eq:AQ_riccati_strategy}), and $\strategies_{\params} = (\strategy^i_{\params})_i$.
% Then it follows that $\tilde{\cstrategies}^{\star} = (\params^{\star},\strategies_{\params})$ is also an SPNE to the game.
% \end{proposition}
% Although the Riccati strategy is the unique linear-feedback equilibrium to the AQ subgame, it is not obviously the unique equilibrium in the general case (\textcolor{red}{cite}).
% We need not concern ourselves with this case, though.

\newcommand{\qedsymbol}{\hfill $\square$}
\begin{proof}
    Clearly, (\ref{eq:config_nash.stage2}) follows for all $i, \params$ from Thm.~\ref{thm:basar.AQ_riccatti}.
    We need only concern ourselves with demonstrating (\ref{eq:config_nash.stage1}), then.
    We proceed by contraposition, assuming there exists $\param^i, \strategy^i$ for some $i \in [N]$ such that
    \begin{equation}
        J^i(\strategy^i(\cdot \,;\, \tilde{\params}),\strategies^{-i\star}_{\tilde{\params}}) < J^i(\strategies^{\star}_{\params^{\star}} \; | \; \params^{\star}) = J^i(\params^{\star})
    \end{equation}
    where $\tilde{\params} = (\param^i,\params^{-i \star})$.
    We must have $\param^i \neq \param^{i \star}$, since otherwise we would have a special case of (\ref{eq:config_nash.stage2}) discussed above, i.e. there is no $\strategy^i(\cdot, \params^{\star}) \neq \strategy^{i\star}_{\params^{\star}}$ achieving a lower cost.
    This logic also holds for $\tilde{\params}$, and so we have no increase in cost if P$i$ plays the Riccati strategy:
    \begin{equation}
        J^i(\strategy^i(\cdot \,;\, \tilde{\params}),\strategies^{-i\star}_{\tilde{\params}}) 
        \geq
        J^i(\strategy^{i\star}_{\tilde{\params}},\strategies^{-i\star}_{\tilde{\params}})  = J^i(\tilde{\params})
    \end{equation}
    Thus, we have $J^i(\tilde{\params}) < J^i(\params^{\star})$, which implies that (\ref{eq:param_cost_Nash}) fails. \qedsymbol
\end{proof}

In general, we do not expect each player's optimization problem $\min_{\param^i} J^i(\param^i, \params^{-i})$ to be convex, and have an example in our experiments where the opposite is true (Sec.~\ref{subsec:general_sum_form}).
Thus, employing gradient-based methods are not gauranteed to get us to global solutions to the single-player problem, let alone the Nash problem.
Instead, we will make due with finding local Nash solutions. as defined here.

\begin{definition}
    \label{defn:local_nash}
    Let an AQ game of configuration be given, satisfying Asm.~\ref{assumption.game_config}.
    We say the configuration strategies $\cstrategies^{
    \star
    } = (\params^{\star},\gamma^{\star})$, with $\params^{\star} \in \Params$ and each $\gamma^{i\star}$ mapping to the Riccati solutions as in Prop.~\ref{prop:stage1_nash}, is a local SPNE to the AQ configuration game if there exists a relatively open neighborhood $U \subseteq \Params, \params^{\star} \in U,$ such that for every $i \in [N]$,
    \begin{equation}
        \label{eq:param_cost_local_Nash}
        J^i(\params^{\star}) \leq J^i(\param^i, \params^{-i \star}), \quad \forall \param^i, (\param^i, \params^{-i \star}) \in U.
    \end{equation}
\end{definition}

The game of configuration, then, reduces to every player optimizing their cost simultaneously.
If each $J^i$ is differentiable and its derivatives reasonably computable, we may approach the game of configuration as a smooth static game over $\Params$, utilizing standard gradient-based approaches.
Given the affine-quadratic setting of this work, we will be able to do so with little difficulty.
For more complex dynamic games in Stage 2, $J^i$ and its gradients will need to be approximated.

 \section{APPROACH AND ANALYSIS}
\label{sec:approach}

In the previous section, we introduced the notion of games of configuration over the space of affine-quadratic scenarios.
We concluded by defining and relating various solution concepts to the game, including both global and local SPNEs.
In this section, we develop a method for searching for such local SPNEs.
Moreover, in doing so, we demosntrate the differentiability of our Stage 1 decision landscape, and derive illuminating first-order necessary conditions for the equilibrium.

\subsection{Gradient of the Value Function}

\begin{theorem}
    \label{thm:gradient}
    Let an AQ game of configuration be given, satisfying Asm.~\ref{assumption.game_config}.
    Then, for every $t \in [0,T]$ and $i \in [N]$, the solutions to the Riccati equation $P^i(\cdot \,;\, \params)$ are differentiable  at every $\params \in \Params$, as are $\zeta^i(\cdot \,;\, \params)$ and $\eta^i(\cdot \, ; \, \params)$.
    The directional derivatives of $P^i(t \, ; \, \params)$ in direction $\bm{h},$ for $\params+\bm{h} \in \Params$, denoted $P^i_{\params; \bm{h}} \triangleq D_{\params} P^i [\bm{h}]$ for brevity, are found as the solutions to the coupled linear equations, provided in \nameref{sec:appendix}.
    % \begin{equation}
    %     \text{\textcolor{red}{(INSERT)}}
    %     % 0= \dot{P}^i_{\params ; \bm{h}} + P^i_{\params ; \bm{h}} \tilde{F} + \tilde{F}^{\top} P^i_{\params ; \bm{h}} + \sum_{j \neq i} P^j_{\params ; \bm{h}} \tilde{H}^{ij} + \tilde{H}^{ij^\top} P^j_{\params ; \bm{h}} + Q^i_{\params ; \bm{h}} + \tilde{G}^{ik}(\params: \bm{h}) \text{\color{red}(Confirm last terms)}
    % \end{equation}
    In particular, the total derivatives of $P^i(t\,; \params), \zeta^i(\cdot), \eta^i(\cdot)$ with respect to $\param^k$, denoted $P^i_{\param^k}, \zeta^i_{\param^k}, \eta^i_{\param^k},$ are given by the linear equations:
    \begin{gather}
        \nonumber 0 = \dot{P}^i_{\param^k} + P^i_{\param^k} \tilde{F} + \tilde{F}^\top P^i_{\param^k} 
        + \sum_{j \neq i} P^{j}_{\theta^k} \tilde{H}^{ij} 
        + \tilde{H}^{ij\top} P^{j}_{\theta^k} \\
        + \left( Q^i_{\theta^k} + \tilde{G}^{ik} \right) ; \quad
        {P}^i_{\theta^i}(T) = 0 \label{eq:P_theta}\\
        \dot{\zeta}^i_{\theta^k} + \tilde{F}^\top \zeta^i_{\theta^k} + \sum_{j \neq i} \tilde{H}^{ij \top} \zeta^j_{\param^k} + \tilde{F}^\top_{\param^k} \zeta^i + W^{ik};
        \quad \zeta^i_{\param^k}(T) = 0, \\
        \nonumber \dot{\eta}^i_{\theta^k} + \beta_{\theta^k}^\top \zeta^i + \beta^\top \zeta_{\theta^k}^i \\
        + \frac{1}{2} \sum_{j=1}^N 2 \zeta^j S^{ij} \zeta^j_{\theta^k} + \zeta^j S^{ij}_{\theta^k} \zeta^j = 0; \quad \eta^i(T) = 0,
        \intertext{where}
        \tilde{H}^{ij} = S^{ij} P^j - S^{jj} P^i \\
        \label{eq:tildeG}
        \tilde{G}^{ik} = P^k S^{ik}_{\param^k} P^k
        - 2 P^i S^{kk}_{\theta^k} P^k \\
        W^{ik} = (P^k S^{ik}_{\param^k} - P^i S^{kk}_{\param^k})  \zeta^k
         + P^i_{\param^k} \beta  + \sum_{j=1}^N P^j_{\param^k} S^{ij} \zeta^j
        % \tilde{G}^{ik} = P^k S^{ik}_{\theta^k} P^k
    % - Z^i S^{kk}_{\theta^k} Z^k - Z^k S^{kk}_{\theta^k} Z^i 
    % \intertext{where}
    % \tilde{H}^{ij} = S^{ij} Z^j - S^{jj} Z^i \\
    % \tilde{G}^{ik} = Z^k S^{ik}_{\theta^k} Z^k
    % - Z^i S^{kk}_{\theta^k} Z^k - Z^k S^{kk}_{\theta^k} Z^i 
    \end{gather}
\end{theorem}
Our expressions for these derivatives can be found formally, by applying standard operations of matrix calculus.
Our proof for the differentiability of $P^i, \zeta^i, \eta^i$ with respect to $\params \in \Params$ requires more care, and so we relegate it to the \nameref{sec:appendix}.
The differentiability of the Stage 1 cost from (\ref{eq:param_cost}) follows immediately from (\ref{eq:AQ_riccati_cost}).
\begin{corollary}
    \label{cor:gradient_of_cost}
    Let an AQ game of configuration satisfying Asm.~\ref{assumption.game_config} be given.
    Then for every $i \in [N]$ and initial condition $x_0 \in \mathbb{R}^n$, the value function $J^i(\params)$ from (\ref{eq:param_cost}) is differentiable with respect to $\params \in \Param$.
    Moreover, the gradient $\nabla_{\params} J^i(\params) = (\frac{d}{d \param^k} J^i(\params) )_k$ is given by the components:
    \begin{equation}
        \label{eq:dJ_dparam}
       \frac{d}{d \param^k} J^i(\params) = \frac{1}{2} x_0^{\top} P^i_{\param^k}(0 \,;\, \params) x_0 + \zeta^i_{\param^k}(0 \,;\, \params)^\top x_0 + \eta^i_{\theta^k}(0 \,;\, \params) .
    \end{equation}
\end{corollary}

Given this differentiability of the cost $J^i$ over $\Params$, local optimality in the sense of Def.~\ref{defn:local_nash} gives rise to the standard kind of first-order, necessary conditions.
\begin{corollary}
    \label{cor:simple_KKT}
    Let $\cstrategies^{\star} = (\params^{\star}, \strategies^\star)$ be a local SPNE to an AQ configuration game satisfying Asm.~\ref{assumption.game_config}, as in Def.~\ref{defn:local_nash}.
    Then for every $i \in [N]$, either $\param^{i\star} \in \partial \Param^i$, i.e. is an endpoint, or $\frac{d}{d \param^i} J^i(\params^{\star}) = 0$, as given in (\ref{eq:dJ_dparam}).
\end{corollary}
These are a very special case of the Karush-Kuhn-Tucker (KKT) conditions, as applied to each player's local optimality in the sense of (\ref{eq:param_cost_local_Nash}).
For the extension higher dimensional spaces for $\Param^i$, one would write these conditions in the standard KKT form with complimentarity conditions enforcing the lower-dimensional boundaries \cite[Thm. 12.1]{nocedal1999numerical}.

Corollary~\ref{cor:simple_KKT} does not provide much insight into the nature of stationary points in configuration games.
By relying on the gradient from Cor.~\ref{cor:gradient_of_cost}, which itself relies on the solution to linear equations (\ref{thm:gradient}), where we might expect $\frac{d}{d\param^i} J^i = 0$ is not immediately obvious.
However, the following expression for this derivative gives us a more meaningful understanding of such critical points.

\begin{proposition}
    \label{prop:envelope_theorem}
    Consider an LQ game of configuration that satisfies Asm.~\ref{assumption.game_config}, and let $\cstrategies^{\star} = (\params^{\star}, \strategies^\star)$ be a fixed strategy, with $\strategies^{\star}$ mapping to the Riccati strategies for each $\params \in \Params$.
    For fixed initial condition $x(0) = x_0$ and ego player $i \in [N]$, let $x^\star(t), t \in [0,T]$ be the optimal trajectory generated by the strategies $\strategies^{\star}(\cdot \,;\, \params^{\star})$,
    % $u^{i\star}(t) = \strategy^i(t, x^{\star}(t) \,\;\, \params^{\star})$
    $u^{i\star}(t)$ the (fixed) open loop representation of their control for $\params^{\star}$, and $u^{j\star}(t,\params) = \strategy^{j\star}(t, x^{\star}(t)\,\;\, \params^{\star})$ the controls for arbitrary $\params$ over $x^\star$ for $j \neq i$.
    Then, letting $\partial_{\param^i}$ denote the partial differentiation operator and omitting $t$ as an argument, we may write the total derivative of the value function as:
    % local SPNE as in Def.~\ref{defn:local_nash}.
    %%
    \begin{gather}
        \nonumber \frac{d}{d \param^i} J^i(\params) = \frac{1}{2} \int_{0}^T
        \left\| x^{\star} \right\|^2_{Q^i_{\param^i}} + 2 x^{\star \top} P^i B^i_{\param^i} u^{i\star} \\
        \label{eq:envelope_theorem}
        + \sum_{j \neq i} \left( \partial_{\param^i} \left\| u^{j \star}(\params^\star) \right\|^2_{R^{ij}} \right) + 2 x^{\star \top} P^i \left( \partial_{\param^i} u^{j \star}(\params^\star) \right) dt
    \end{gather}
    In particular, if $\cstrategies^\star$ is a local SPNE, either $\param^{i\star} \in \partial \Param^i$ or this integral is zero.
\end{proposition}
The terms in this integral illuminate which trade-offs in parameter selection must be balanced in order for each player to be at equilibrium.
Each term is readily interpretable:
\begin{itemize}
    \item The first term, $\left\| x^{\star} \right\|_{Q^i_{\param^i}}^2$, is the impact of further tuning $\param^i$ on the instantaneous state cost.
    \item The second term $x^{\star \top} P^i B^i_{\param^i} u^{i\star}$ tracks how much more (or less) effective P$i$'s control input would become, with respect to optimizing the (P$i$) value function.
    \item The first summand tracks any incurred control cost \textbf{due to P$j$'s strategic shift}.
    \item The second summand tracks the instantanous impact of \textbf{P$j$'s strategic shift} to P$i$'s value function.
\end{itemize}
Notably, and perhaps somewhat surprisingly, there is no accounting for indirect impact on (\ref{eq:envelope_theorem}) from strategic interactions on P$i$'s behalf, i.e. through $\partial_{\theta^i} u^i$, while \textbf{there is such impact from other players' shifted strategies, only}. 
In fact, Proposition~\ref{prop:envelope_theorem} is a special case of the envelope theorem for finite-horizon differential games \cite{caputo2007envelope}, which states that this is a general property.
We prove it directly in the \nameref{sec:appendix}.

\begin{algorithm}
\caption{Iterated Best Response for AQ Game of Configuration}
\label{alg:solve_two_stage}
\begin{algorithmic}
\REQUIRE Game $\mathfrak{G}(\params)$, initial condition $x_0$, initial guess $\params_0 \in \Params$, step size $\alpha \in \mathbb{R}$, tolerance $\epsilon > 0$, max iterations $\ell_{\max}, \tau_{\max}$
\STATE $\ell \gets 0,\; \params_{\ell} \gets \params_0$
\WHILE{ $\lVert \params_{\ell} - \params_{\ell-1} \rVert > \epsilon$ \AND $\ell < \ell_{\max}$ }
    \STATE $\ell \gets \ell + 1,\; \tau \gets 0$
    \FOR{ $i \in [N]$ }
        \STATE $\tilde{\params}_{0} \gets (\params^{1:i-1}_{\ell},\params^{i:N}_{\ell-1}),\; \tau \gets 0$
        \WHILE{ $\lVert \param^i_{\tau} - \param^i_{\tau-1} \rVert > \epsilon$ \AND $\ell < \ell_{\max}$ }
            \STATE $\tau \gets \tau + 1$
            \STATE $\param^i_{\tau} \gets \param^i_{\tau-1} - \alpha \frac{dJ^i}{d\param^i}(\tilde{\param}^i_{\tau-1}, \tilde{\params}^{-i}_{0})$
            \STATE $\param^i_{\tau} \gets \mathrm{project}(\param^i_{\tau})$ \COMMENT{Onto $\Param^i$}
        \ENDWHILE
        \STATE $\param^i_{\ell} \gets \tilde{\param}^i_{\tau},\; \tau \gets \tau + 1$
    \ENDFOR
\ENDWHILE
\RETURN $\params_{\ell},\; \mathrm{converged} \gets \mathrm{bool}(\lVert \params_{\ell} - \params_{\ell-1} \rVert \leq \epsilon)$
\end{algorithmic}
\end{algorithm}

These necessary conditions shed light on the behavior we would expect to candidate equilibria.
We incorporate these gradients into a standard iterated best response (IBR) method for searching for local equilibria \cite{wang2019game}.
% \textcolor{red}{(CITE)}
An overview of our algorithm is given in Alg.~\ref{alg:solve_two_stage}.
Besides the computation of the gradients, the method we are employing is otherwise standard.

% \begin{algorithm}[ht]
% % \SetAlgoLined
% % \DontPrintSemicolon
% \caption{Iterated Best Response for AQ Game of Configuration}
% \label{alg:solve_two_stage}
% {
% \small
%     % \textbf{Define:} Parametrized game $\mathfrak{G}(\params) = (A(t), B^i(t; \param^i) Q^i(t;\params), R^{ij}(t),)$ \\
%     \textbf{Input:} Game $\mathfrak{G}(\params)$, initial condition $x_0$, initial guess $\params_0 \in \Params$, step size $\alpha \in \mathbb{R}$, tolerance $\epsilon > 0$, max iterations $\ell_{\max}$.\\
%     % $\structs \gets \structs_0$\\
%     \While{$!a$}{
%        $a+b$\\
%        $c+d$
%     % $\structs_{0} \gets \structs - \alpha \frac{d\structcost}{d\structs}$\\
%     % $\structs \gets \mathrm{project}(\structs_{0})$ \tcp{Onto constraints}
%     }
% $\var^1, \var^2 \gets \mathrm{solve Stage 2}(\structs, p(\world))$\\
%     \Return{$\structs,\var^1, \var^2$}
% }
% \end{algorithm}

\begin{figure}[t]
    \centering
    \includegraphics[scale=0.28]{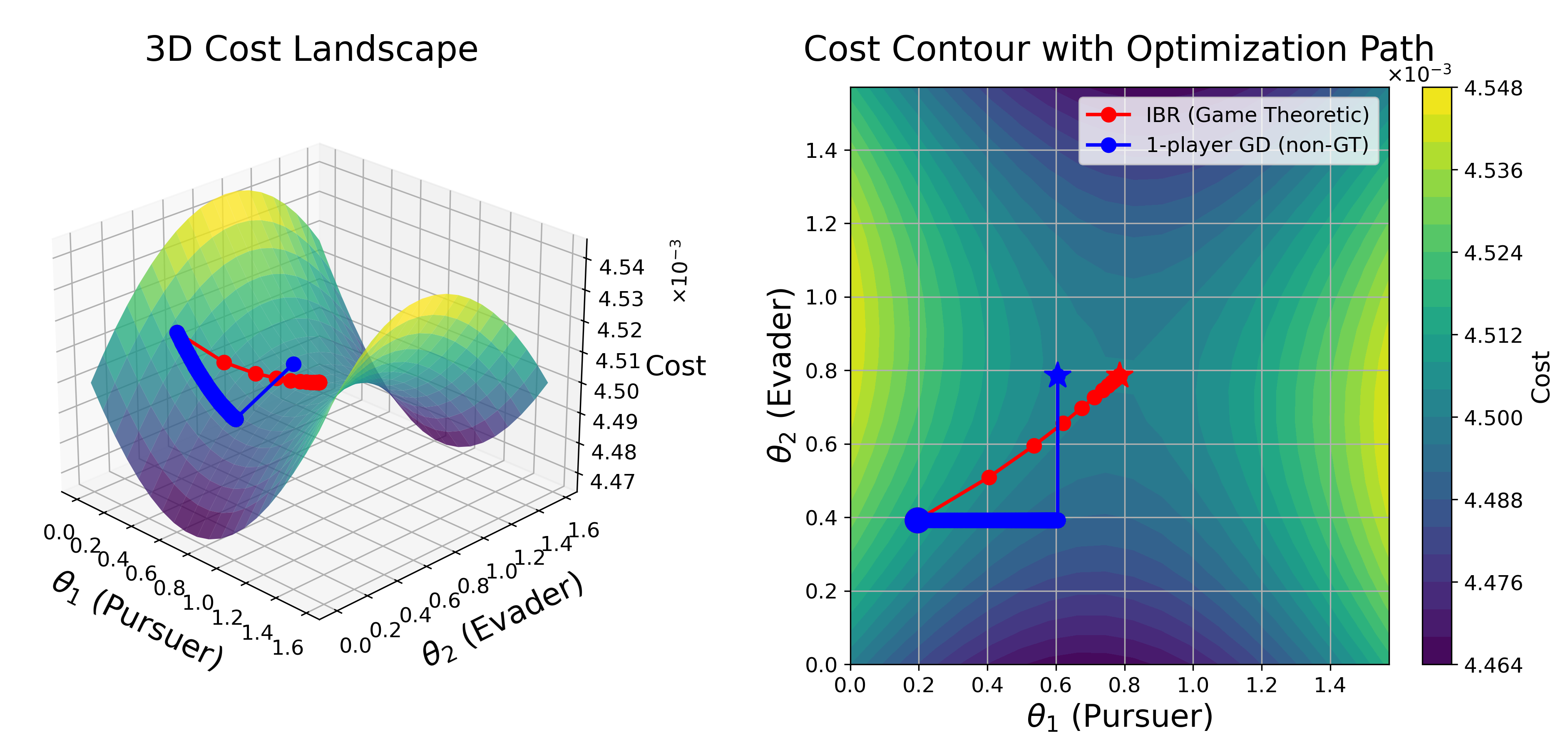}
    \caption{\textbf{Finding configuration saddle-point for LQ pursuit-evasion game.} Using our IBR-based method (Alg.~\ref{alg:solve_two_stage}), we are able to find the saddle-point equilibrium over configurations for a finite-horizon pursuit-evasion game (red).
    For comparison, we also consider the scenario with a non-game-theoretic pursuer, who aims for optimality against the initial evader configuration instead of the Nash (blue).
    % As a point of comparison, we consider the scenario where only the evader anticipates mutual (adversarial) configuration, while the pursuer only fine-tunes their own parameter while falsely assuming the static, initial configuration on the part of their adversary (BLUE).
    % As we can see, in that scenario the better configured evader is able to gain more distance/cost on the pursuer.
    }
    \label{fig:PEgame}
\end{figure}

\begin{figure*}[t]
    \centering
    \includegraphics[scale=0.5]{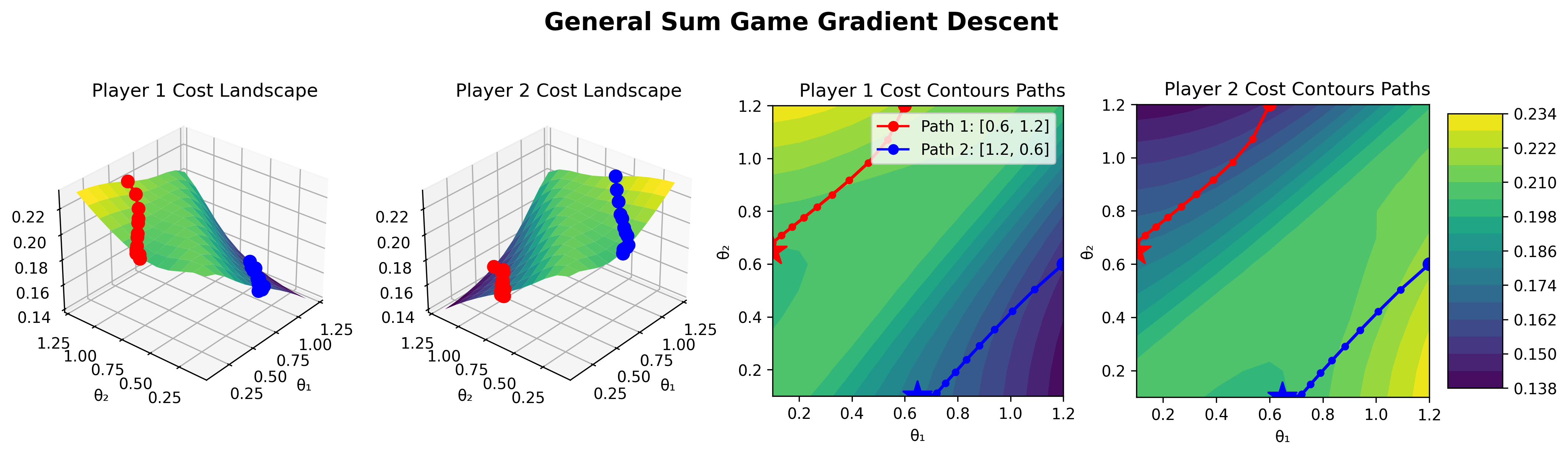}
    \caption{This is our Gradient Descent for a challenging General Sum Case for both players. We set our parameter as $Q_v=25$, $w_r=0.02$ and $Q_h =100*(0.5\text{sign}(3-t)+0.5)$ based on Eq.~\ref{eq:general_cost_matrix}. $\text{sign}(t)$ represents the sign function which is $1$ for $t\geq0$ and $-1$ for $t<0$. Initial conditions, $[\theta^1,\theta^2]$, are $[0.6,1.2]$ for the red path and  $[1.2,0.6]$ for the blue.
    \label{fig:General_sum_game}}
\end{figure*}

\section{EXPERIMENTS}
\label{sec:experiments}

In the previous section, we developed a method for computing the gradients of the configuration costs for a finite-horizon AQ game, envisioning the task of multi-agent configuration as the first in a two-stage scenario.
We used these to formulate an IBR method for finding local solutions.
%% \jesse{Jeff, I will work this text in}
% \jeff{For our following section, we design our experiments to show that first without adding a game theoretic approach the results are sub-optimal for both a general sum and zero sum example. Next, the addition  game theory creates more optimal solutions for the agent.
% Our zero sum game is the pursuit evasions game where one agent is trying to catch the other. Before the game, the agents can choose wether to have greater maneuverability such as changing directions or velocity. 
% In the case where only one of the agents considers the game theory elements, that agent is favored to win the game.
% We also further demonstrate that if both players consider game theory elements the solution is logically consistent with the intuition. 
% Our general sum game is two trains on tracks have to both cross an intersection and reach an optimal velocity forward. Before the game begins, the can personally choose how much to penalize their acceleration in the quadratic costs.
% If the trains do not consider each other as agents capable of decisions, they naively choose the minimal possible penalty for their acceleration and collide. 
% However if we add a game theoretic formulation to this problem, the trains are able to successfully negotiate the intersection and avoid a collision by having one agent choosing a higher penalty arriving at the interesction later and another chooses a smaller penalty.}
In this section, we provide example games of configuration, and demonstrate that our method is able to reliably find local solutions. 

% \textcolor{red}{TO DO}
% \begin{itemize}
%     \color{red}
%     \item finish intro
%     \item zero-sum, pursuit-evasion example with saddle point
%     \item train intersection system, or formation control as fallback
%     \item maybe analysis of robustness of approach? Comparison of different methods? Let's see how much room we have
% \end{itemize}

\subsection{Zero-Sum Pursuit Evasion Game Formulation}
\label{subsec:PE_game}

We begin by considering a zero-sum pursuit-evasion game.
In such a setting, $J \triangleq J^1 = - J^2$ and thus any alternating gradient-based approach boils down to a search for local saddle-point equilibrium on the same landscape for both players.

We consider the two-player linear-quadratic system given by (\ref{eq:LQ.dynamics}, \ref{eq:LQ.cost}), with state vector $x = (x^1, x^2)$, and where each player has private state $x^i = (p_x^i, p_y^i, v_x^i, v_y^i) \in \mathbb{R}^4$.
The intrinsic dynamics of the system are stationary, and given by the $8\times 8$ block-diagonal matrix $A = \diag(A', A')$, where $A'$ is given in block form as
\begin{equation}
    A' = \begin{bmatrix}
        0_2 & I_2 \\
        0_2 & 0_2 
    \end{bmatrix},
\end{equation}
P$1$'s $8 \times 2$ control matrix $B^1(\param^1)$ is given as
\begin{gather}
    B^1(\param^1) = \kappa_1 [0_{2 \times 2}, \quad B'(\param^1), \quad 0_{2 \times 4}]^\top,\\ B'(\param) = \diag( 1 + \cos \theta, 1 + \sin \theta),
\end{gather}
for a scalar $\kappa_1 \in (0,1)$, and P$2$'s is analogously defined with scalar $\kappa_2$.
Each player will be choosing a parameter $\param^i \in \Param^i = [0, \pi/2]$.
For the cost matrices, we have $Q=0, R^i = I$ and the terminal cost:
\begin{equation}\label{eqn:collision_gain}
    Q_f = \kappa_3 \begin{bmatrix}
        I_2 & 0_{2} & -I_2 & 0_{2} \\
        0_{2} &  0_{2} & 0_{2} & 0_{2}\\
        -I_2 & 0_{2} & I_2 & 0_{2} \\
        0_{2} &  0_{2} & 0_{2} & 0_{2}
    \end{bmatrix}
\end{equation}
In general, the Riccati equation for this system is not stable, and thus there is a sufficiently large $T>0$ for which the ODE given by (\ref{eq:riccati_zerosum}) becomes unbounded.
However, by appropriately tuning $\kappa_1, \kappa_2, \kappa_3$, we can find finite-horizon solutions. We consider $\kappa_1=\kappa_2 = 1$ and $\kappa_3 = 0.0005$.

From this parametrized LQ game, we form a configuration game by allowing both players to choose their parameter $\param^i \in \Param^i = [0, \pi/2]$, which determines their favored direction of travel.
Choosing $\param$ close to zero allows the agent to more easily travel in the horizontal direction, while $\param$ closer to $\pi/2$ allows them to travel vertically.
Starting from an initial state $\params_0$, we proceed using our gradient-based IBR method (Alg.~\ref{alg:solve_two_stage}) until the algorithm converges to a local solution $\params_{\ell}$ after $\ell$ timesteps.
For this particular system, the Stage 1 cost landscape over $\Params = [0, \pi/2]^2$ turns out to form an apparently concave-convex landscape with a unique saddle-point equilibrium.
Thus, our algorithm easily finds the global solution (Fig.~\ref{fig:PEgame}, red path).
As a point of comparison, we also consider the scenario in which only the evader, and not the pursuer, is anticipating adversarial configuration.
Here, the evader chooses their saddle-point value for $\theta^{2\star}$, while the pursuer merely performs gradient descent for $\param^1$ from the initial $\params_0$, arriving at $\tilde{\param}_\tau^1 = \arg \min_{\param^1} J(\param_\tau^1, \param_0^2)$.
The realized value of the game is then given by $J(\tilde{\param}_{\tau}^1,\param^{2\star})$.
As we can see in Fig.~\ref{fig:PEgame}, blue path, this approach leads to a suboptimal solution for the pursuer.

\subsection{General-Sum  Game Results}\label{subsec:general_sum_form}
Additionally, we can apply our methodology to  general sum games.
We pick a $1$-D motion system defined as follows:
\begin{equation}
    \dot{\mathbf{x}}= \begin{bmatrix}
     0& 1  &0&0\\
     0&0&0 &0\\
     0 &0 & 0 &1\\
     0 &0& 0 &0
    \end{bmatrix}\begin{bmatrix}
    p^1 \\
    \dot{p}^1 \\
    p^2 \\
    \dot{p}^2
    \end{bmatrix}  +  \begin{bmatrix}
    0\\
    \theta^1\\
     0\\
    0
    \end{bmatrix}\mathbf{u}^1 +  \begin{bmatrix}
    0\\
     0\\
     0 \\
          \theta^2
    \end{bmatrix}\mathbf{u}^2
\end{equation}

$p^i$ and $u^i$ is agent's $i$ position and control input respectively. $\theta^i$ is the tunable parameter representing an agent's aggression with respect to acceleration. The objective is

    \begin{equation}\label{eqn:general_cost}
\begin{aligned}
    \min_{\mathbf{u}} \quad & \frac{1}{2} (\mathbf{x} - \mathbf{f})^{\top} \mathbf{Q} (\mathbf{x} - \mathbf{f}) + \mathbf{u}^\top\mathbf{u} + w_r\exp(-10 (\theta^1 - \theta^2)^2) \\
    \quad & \mathbf{f} = \begin{bmatrix} 0 & v^1_o & 0 & v^2_o \end{bmatrix}^\top.
\end{aligned}
\end{equation}

  First, let $Q_v \in \mathbb{R}$ be the cost for tracking an optimal forward velocity $v^i_0$ and 
$Q_h(t) \in \mathbb{R}$ represents a time-varying penalty for the term $(p^2-p^1)^2$. We then define $\mathbf{Q}$ matrix as

\begin{equation}\label{eq:general_cost_matrix}
    \mathbf{Q} = \begin{bmatrix}
        -Q_h(t) & 0  & Q_h(t) & 0 \\
        0 & Q_v & 0 & 0 \\
        Q_h(t) & 0 & -Q_h(t) & 0 \\
        0 & 0 & 0 & Q_v
    \end{bmatrix}.
\end{equation}

While not an exact real world model, this objective has many properties analogous to real world scenarios. 
For example, imagine two parallel trains wanting to maintain their desired speed, but not wanting to arrive at their destinations at the same time so as to avoid congestion.
Alternatively this could represent two cars wanting to maintain distance and their desired speed when merging lanes.
As seen in Fig.~(\ref{fig:General_sum_game}), P$1$ desires the parameter regime $[1.2,0.2]$ and P$2$ wants the reverse $[0.2,1.2]$. Physically, in the lane merging example, this might mean neither car would want to slow down to create this distance but hopes the other will slow instead. 
Additionally, we added a highly nonlinear exponential term in the regularizer and show a time varying $Q_h(t)$ parameter regime, merely to complicate the optimization  landscape.
Despite these challenging conditions, as we see in Fig.~(\ref{fig:General_sum_game}) our method is able to converge to local minima with both competing desires and a nonlinear regularizer.

\section{CONCLUSIONS}

We present a novel framework for non-cooperative, multi-agent configuration for dynamic scenarios.
We analyzed the gradients of these decisions and used them to find local solutions to configuration games in the affine-quadratic setting.
This was the natural first step for differential games, but such systems are highly limited in their real-world modeling capability.
In our future work, we seek to extend this approach to the iterative LQ setting (as in \cite{oshin2022parameterized} and \cite{fridovich2020efficient}, for instance), in order to apply our approach to more general, real-world multi-agent systems.

%%%%%%%%%%%%%%%%%%%%%%%%%%%%%%%%%%%%%%%%%%%%%%%%%%%%%%%%%%%%%%%%%%%%%%%%%%%%%%%%

%%%%%%%%%%%%%%%%%%%%%%%%%%%%%%%%%%%%%%%%%%%%%%%%%%%%%%%%%%%%%%%%%%%%%%%%%%%%%%%%

%%%%%%%%%%%%%%%%%%%%%%%%%%%%%%%%%%%%%%%%%%%%%%%%%%%%%%%%%%%%%%%%%%%%%%%%%%%%%%%%

\section*{APPENDIX}\label{sec:appendix}

In this section, we will need the following fact. Suppose we have a linear differential equation of the form
$   \dot{M} + \tilde{A}^\top M + M A + W; \quad M(T) = 0
$
with all terms continuous on $[0,T]$, but potentially time-varying.
Then the solution $M(t), t\in [0,T]$ may be written as the integral equation:
\begin{equation}
    \label{eq:fundamental_matrix_integral}
    M(t) = \Phi(T,t)^\top M_T \Phi(T,t) + \int_{t}^T \Phi(s,t)^\top W(s) \Phi(s,t) ds.
\end{equation}
Proposition \ref{prop:envelope_theorem} follows relatively straight-forwardly from applying this formula to (\ref{eq:P_theta}) in the LQ case, and then using the substitution $S^{jj}P^j x^{\star} = B^j u^{j\star}$ and its derivatives where appropriate. We omit the proof.

We first provide an abbreviated version of our proof for the main result.
Many details had to be omitted, but the authors are more than happy to provide any missing details.
All matrix norms are in the sense of Frobenius, i.e. $\left\| K \right\| = \sqrt{\sum_{ij} k_{ij}^2}$.

\textbf{Proof of Theorem \ref{thm:gradient}.}

\begin{proof}
Let $\params \in \Params$ be given, and direction $\bm{h}$ such that $\params + \bm{h} \in \Params$, where we take $\| \bm{h} \|_2 = 1$ without any loss of generality.
We will begin by proving the existence of the directional derivatives of $P^i(t \,;\, \params)$, for every $i$, with respect to $\params$ in the direction $\epsilon \bm{h}$ for arbitrary $\epsilon \in (0,1)$. i.e $D_{\params} P^i (\epsilon \bm{h})$.
We employ the simplifying notation $P^i_{\params; \epsilon h} \triangleq D_{\params} P^i (\epsilon \bm{h})$, and similarly for other such matrix derivatives.
Note that we omit the argument $t$ when it is fixed or clear from context.
We will prove that $P^i_{\params; \epsilon \bm{h}}$ exist for all $i, t$, and are found as the solutions to the coupled linear ODEs:
\begin{gather}
    \nonumber
    \dot{P}^i_{\params; \epsilon \bm{h}} + \tilde{F}^\top P^i_{\params; \epsilon \bm{h}} + P^i_{\params; \epsilon \bm{h}} \tilde{F}
    + \sum_{j \neq i} \left( P^i_{\params; \epsilon \bm{h}} H^{ij}  +   H^{ij\top} P^i_{\params; \epsilon \bm{h}} \right) \\
    + Q^i_{\params; \epsilon \bm{h}} + G^{i}(\params ; \epsilon \bm{h}); \quad P^i_{\params; \epsilon \bm{h}}(T) = 0
    \label{eq:appendix_total_deriv}
\end{gather}
where $\tilde{H}^{ij}$ as in (\ref{eq:tildeG}) and $ G^{i}(\params ; \epsilon \bm{h}) = \sum_{j} P^j S^{ij}_{\params; \epsilon \bm{h}}P^j - P^i S^{jj}_{\params; \epsilon \bm{h}} P^j - P^j S^{jj}_{\params; \epsilon \bm{h}} P^i.$
It is straight-forward to prove that this may be rewritten, in concatenated form, as $\dot{\bm{P}}_{\params; \epsilon \bm{h}} +  \mathcal{L}_{\params}(\bm{P}_{\params; \epsilon \bm{h}}; \epsilon \bm{h}) = 0$, where $\Lambda_{\params}$ is bilinear in both arguments. We will use this later.
By the conditions of Asm.~\ref{assumption.game_config}, we have that $Q^i$ and $S^{ij}$ are differentiable, and their expansions are given by
\begin{gather}
     Q^i(\params + \epsilon \bm{h}) - Q^i(\params) = \epsilon D_{\params} Q^i \bm{h} + \epsilon^2 M_{Q^i}\\
     S^{ij}(\params + \epsilon \bm{h}) - S^{ij}(\params) = \epsilon D_{\params} S^{ij} \bm{h} + \epsilon^2 M_{S^{ij}}
\end{gather}
where the remainder terms $M_{Q^i}, M_{S^{ij}}$ are uniformly bounded above 
$\left\| 
M_{Q^i} \right\|_{\mathcal{F}}, \left\| 
M_{S^{ij}} \right\|_{\mathcal{F}} \leq \beta$ for some $\beta$, for all $i,j$ and all $t, \bm{h}, \epsilon$.
This will let us control higher order terms.
Moreover, via their continuity on $[0,T] \times \Params$, we can take $\sup_{i,j, \params, t}(S^{ij},\left\| Q^i \right\|) \leq \beta$ as well.
Finally, without loss of generality, the Lipschitz continuity of all $P^i$ on $[0,T]\times \Params$ implies that $\left\| P^i(\params) - P^i(\params') \right\| \leq \ell \left\| \params - \params' \right\|$ and $\left\| P^i( \params) \right\| \leq \ell$ as well, for all $i, t, \params, \params'$.

Let $\Delta^i_\epsilon = \Delta^i_{\params, \epsilon \bm{h}} \triangleq \frac{1}{\epsilon}(P^i(\params + \epsilon \bm{h}) - P^i(\params))$ for all $i, t$.
Observe that $\Delta^i_{\epsilon}$ is thus defined and differential over time on $[0,T]$ as a scaled sum of such functions, and moreover $\left\| \Delta^i_{\epsilon} \right\| \leq \epsilon \ell \left| \bm{h} \right\|_2/\epsilon = \ell$ via Lipschitz continuity of $P^i$.
Then we may write its ODE by combining those of $P^i(\params)$ and $P^i(\params + \epsilon \bm{h})$, exploiting the expansions of $Q^i$ and $S^{ij}$ above about $\params$, and arrive at
\begin{gather}
    % \dot{\Delta}^i_{\epsilon \bm{h}} = \Delta \tilde{F} + \tilde{F}^\top \Delta - () + \left[ Q^i_{\params; \bm{h}} \bm{h} + \epsilon M_{Q^i} \right] \\
    % + \sum_{j} P^j S^{ij}_{\params, \bm{h}} P^j + 2 \Delta^j_{\epsilon} S^{ij} P^j + \epsilon E^{j}_1; \quad \Delta_h(T) = 0 \\
    \label{eq:appendix_delta_i}
    \dot{\Delta}^i_{\epsilon \bm{h}} + \mathcal{L}^i_{\params}(\dot{\Delta}^i_{\epsilon \bm{h}}; \epsilon \bm{h}) + \epsilon \tilde{E}^i
\end{gather}
where for all $t,i, \epsilon,$ we have the common bounds on the remainder $\tilde{E}^i = \sum_{j} E^{ij}_1 + E^{ij}_2$, where
\begin{gather}
    \left\| E^j_1 \right\|^2 \leq (6 + 6 \epsilon + \epsilon^2) \ell^2 \beta) \\
    \left\| E^j_2 \right\|^2 \leq (4 + 3 \epsilon + 2 \epsilon^2) \ell^2 \beta
\end{gather}
where these remainders, respectively, come from the 2nd/3rd and 4th/5th terms in (\ref{eq:appendix_total_deriv}).
If one uses use the formula from (\ref{eq:fundamental_matrix_integral}) to derive similar equations for the solutions to (\ref{eq:appendix_total_deriv},\ref{eq:appendix_delta_i}), after cancellation one arrives at
\begin{equation}
    \mathcal{D}^i(t) = P^i_{\params; \epsilon \bm{h}} - \Delta^i_{\epsilon \bm{h}} = \epsilon \int_{t}^T \Phi(s,t)^\top \tilde{E}^i(t) \Phi(s,t) ds
\end{equation}
Since $\tilde{E}^i$ is uniformly bounded, we have $\Delta^i_{\epsilon \bm{h}} \to P^i_{\params; \epsilon \bm{h}}$ as $\epsilon \to 0$ as we have defined it.
Moreover, $P^i_{\params; \bm{h}}$ inherits linearity in $\bm{h}$ from $\mathcal{L}_{\params}$, and thus we have differentiability.
\end{proof}

\section*{ACKNOWLEDGMENT}

We would like to thank David Fridovich-Keil for his helpful comments on an early draft of this paper, Xinjie Liu (UT Austin) for discussion of the theory, and Yang Zhou (NYU) for discussions on the appropriate applications of this work.

%%%%%%%%%%%%%%%%%%%%%%%%%%%%%%%%%%%%%%%%%%%%%%%%%%%%%%%%%%%%%%%%%%%%%%%%%%%%%%%%

% References are important to the reader; therefore, each citation must be complete and correct. If at all possible, references should be commonly available publications.

\bibliographystyle{ieeetr}
\bibliography{references}

\begin{thebibliography}{10}

\bibitem{loquercio2022autotune}
A.~Loquercio, A.~Saviolo, and D.~Scaramuzza, ``Autotune: Controller tuning for high-speed flight,'' {\em IEEE Robotics and Automation Letters}, vol.~7, no.~2, pp.~4432--4439, 2022.

\bibitem{marco2016automatic}
A.~Marco, P.~Hennig, J.~Bohg, S.~Schaal, and S.~Trimpe, ``Automatic lqr tuning based on gaussian process global optimization,'' in {\em IEEE international conference on robotics and automation (ICRA)}, pp.~270--277, IEEE, 2016.

\bibitem{cheng2024difftune}
S.~Cheng, M.~Kim, L.~Song, C.~Yang, Y.~Jin, S.~Wang, and N.~Hovakimyan, ``Difftune: Auto-tuning through auto-differentiation,'' {\em IEEE Transactions on Robotics}, 2024.

\bibitem{romero2023weighted}
A.~Romero, S.~Govil, G.~Yilmaz, Y.~Song, and D.~Scaramuzza, ``Weighted maximum likelihood for controller tuning,'' in {\em IEEE International Conference on Robotics and Automation (ICRA)}, pp.~1334--1341, 2023.

\bibitem{configure_riveting}
H.~Girgin, T.~S. Lembono, R.~Cirligeanu, and S.~Calinon, ``Optimization of robot configurations for motion planning in industrial riveting,'' in {\em 2021 20th International Conference on Advanced Robotics (ICAR)}, pp.~247--252, 2021.

\bibitem{oshin2022parameterized}
A.~Oshin, M.~D. Houghton, M.~J. Acheson, I.~M. Gregory, and E.~A. Theodorou, ``Parameterized differential dynamic programming,'' {\em arXiv preprint arXiv:2204.03727}, 2022.

\bibitem{fridovich2020efficient}
D.~Fridovich-Keil, E.~Ratner, L.~Peters, A.~D. Dragan, and C.~J. Tomlin, ``Efficient iterative linear-quadratic approximations for nonlinear multi-player general-sum differential games,'' in {\em IEEE international conference on robotics and automation (ICRA)}, pp.~1475--1481, 2020.

\bibitem{caputo2007envelope}
M.~R. Caputo, ``The envelope theorem for locally differentiable nash equilibria of finite horizon differential games,'' {\em Games and Economic Behavior}, vol.~61, no.~2, pp.~198--224, 2007.

\bibitem{risk_aware_planning_wang}
M.~Wang, N.~Mehr, A.~Gaidon, and M.~Schwager, ``Game-theoretic planning for risk-aware interactive agents,'' in {\em IEEE/RSJ International Conference on Intelligent Robots and Systems (IROS)}, pp.~6998--7005, 2020.

\bibitem{williams2023distributed}
Z.~Williams, J.~Chen, and N.~Mehr, ``Distributed potential ilqr: Scalable game-theoretic trajectory planning for multi-agent interactions,'' in {\em IEEE International Conference on Robotics and Automation (ICRA)}, pp.~01--07, 2023.

\bibitem{GLIZER201522}
V.~Y. Glizer and V.~Turetsky, ``Linear-quadratic pursuit-evasion game with zero-order players’ dynamics and terminal constraint for the evader,'' {\em IFAC-PapersOnLine}, vol.~48, no.~25, pp.~22--27, 2015.
\newblock 16th IFAC Workshop on Control Applications of Optimization CAO’2015.

\bibitem{layout_design_multi_robot_transport}
C.~Bosio and M.~W. Mueller, ``Automated layout and control co-design of robust multi-uav transportation systems,'' {\em IEEE Robotics and Automation Letters}, vol.~10, no.~4, pp.~3956--3963, 2025.

\bibitem{ZHOU2022105376}
L.~Zhou, Y.~Zheng, Q.~Zhao, F.~Xiao, and Y.~Zhang, ``Game-based coordination control of multi-agent systems,'' {\em Systems and Control Letters}, vol.~169, p.~105376, 2022.

\bibitem{WANG20237858}
L.~Wang, T.~Qiu, Z.~Pu, J.~Yi, J.~Zhu, and Y.~Zhao, ``A decision-making method for swarm agents in attack-defense confrontation,'' {\em IFAC-PapersOnLine}, vol.~56, no.~2, pp.~7858--7864, 2023.
\newblock 22nd IFAC World Congress.

\bibitem{fisac2019hierarchical}
J.~F. Fisac, E.~Bronstein, E.~Stefansson, D.~Sadigh, S.~S. Sastry, and A.~D. Dragan, ``Hierarchical game-theoretic planning for autonomous vehicles,'' in {\em IEEE International Conference on Robotics and Automation (ICRA)}, pp.~9590--9596, 2019.

\bibitem{bacsar1998dynamic}
T.~Ba{\c{s}}ar and G.~J. Olsder, {\em Dynamic noncooperative game theory}.
\newblock SIAM, 2~ed., 1998.

\bibitem{nocedal1999numerical}
J.~Nocedal and S.~J. Wright, {\em Numerical optimization}.
\newblock Springer.

\bibitem{wang2019game}
Z.~Wang, R.~Spica, and M.~Schwager, ``Game theoretic motion planning for multi-robot racing,'' in {\em Distributed Autonomous Robotic Systems: The 14th International Symposium}, pp.~225--238, Springer, 2019.

\end{thebibliography}

\end{document}